# Strategies for Imaging Faint Extended Sources in the Near-Infrared

Ovidiu Vaduvescu and Marshall L. McCall

*York University, Department of Physics and Astronomy, 4700 Keele Street, Toronto, ON, Canada, M3J 1P3*

`ovidiuv@yorku.ca`

## ABSTRACT

Quantitative information about variations in the background at $J$ and $K'$ are presented and used to develop guidelines for the acquisition and reduction of ground-based images of faint extended sources in the near-infrared, especially those which occupy a significant fraction of the field of view of a detector or which are located in areas crowded with foreground or background sources. Findings are based primarily upon data acquired over three photometric nights with the $3\!.\!6 \times 3\!.\!6$ CFHT-IR array on the Canada-France-Hawaii Telescope atop Mauna Kea. Although some results are specific to CFHT, overall conclusions should be useful in guiding observing and reduction strategies of extended objects elsewhere. During the run, the mean brightness of the background (more than 70% of which was from the sky) varied significantly on a very short time scale, by 0.7% per minute in $J$ and 0.5% per minute in $K'$, on average. Changes in the optical depth of the sky were partly responsible, because stars faded as the background level increased. A changing pattern in the background was evident from differences of consecutive pairs of frames (0.3% per minute in $J$ and 0.2% per minute in $K'$), but this originated primarily in the instrumentation. Any pattern over $3\!.\!6$ associated with the atmosphere changed at a rate less than about 0.06% per minute in $K'$ relative to the signal from the sky alone. To measure the background to a precision of 1% per frame, exposures of extended targets should be alternated with identical exposures of the background. In $J$ and $K'$, target and sky exposures ought to be separated by no more than 90 s and 130 s, respectively. To observe a target larger than about 40% of the field of view, background samples ought to be taken with the target shifted completely out of the field. For smaller targets, gains in efficiency can be made by shifting the target to a different place on the array. The signal-to-noise ratio of the reduced image of a target is maximized by evaluating the background for each individual image using only the samples taken immediately before and after. Provided background images are dithered, it is possible to recognize and remove celestial contaminants through differencing.

*Subject headings:* infrared: galaxies, infrared: general, galaxies: photometry, methods: statistical, techniques: image processing.

## 1.   Introduction

In any given near-infrared (NIR) image of a celestial body, the background comes from three sources:

1. the atmosphere (e.g., airglow)

2. the detector (dark and bias)

3. thermal radiation from structures above the detector, both filtered and unfiltered

To zeroth order, the background shows up as a large offset in signal across any given frame. The atmospheric component usually predominates in $J$ and $K'$, the brightness of which changes rapidly with time. There may be spatial variations in the background, also, such as from any non-uniformity of the dark current, bias, response, or airglow. Such variations may not be stable temporally, either. Of course, the contribution from thermal radiation changes with time as the temperature varies. Worse still, that part of the thermal radiation from the detector's housing will not have been filtered in the same way as light from above.

Imaging of faint extended sources in the near-infrared (NIR) is made particularly difficult because the atmospheric component of the background is so bright. For example, $J \sim 16 \, \mathrm{mag \, arcsec^{-2}}$ and $K \sim 13 \, \mathrm{mag \, arcsec^{-2}}$ for a dark site like Chile or Mauna Kea (Vanzi & Hainaut 2002). For Mauna Kea, $K' \sim 14 - 15 \, \mathrm{mag \, arcsec^{-2}}$ at unit airmass (Wainscoat & Cowie 1992). For galaxies, the surface brightness of the isophote defining $D_{25}$ in the $B$ band corresponds to 21 to $23 \, \mathrm{mag \, arcsec^{-2}}$ in $K$ (e.g., Fioc & Rocca-Volmerange 1999), which is 8 to 10 mag fainter than the sky!

To carry out surface photometry of a faint extended source, it is critical to define the background under the source precisely. How well the background is defined determines not only the limiting detectable surface brightness but also the shape of the surface brightness profile at surface brightnesses which are a small percentage of the sky. In the NIR, small errors in flat-fielding may have catastrophic consequences for surface photometry because of the overwhelming brightness of the background relative to the source (see Melnick et al. (1999) for an interesting discussion of this point with respect to photometry in $I$).

Consider a target which is large compared to the field of view of a detector, but does not fill it, for which the surface brightness is uniformly $23\,\mathrm{mag\,arcsec^{-2}}$ in $K'$. Assume that the sky brightness is $15\,\mathrm{mag\,arcsec^{-2}}$. Suppose that an image is acquired at $K'$, and that any bias, dark current, and unfiltered thermal radiation are somehow removed to the point that the image can be corrected for variations in response. Upon dividing by a flat field image, suppose that low-frequency variations in response are imperfectly flattened, and that the corners end up being depressed relative to the centre by 0.5% (as might be typical in the optical). If the corners of the image are used to interpolate the background at the centre of the image, the surface brightness of the source at its core will be judged to be 2.4 mag brighter than it actually is, and the surface brightness profile will be judged to be peaking towards the center! On the other hand, if the background across the field could be measured and subtracted to 1% accuracy, then the flattening error would lead to an estimate for the central surface brightness deviating by only 0.1 mag from the actual value.

Thermal radiation from the detector housing is a particularly troublesome aspect of NIR observing. Because it does not pass through the filter, the response pattern superimposed upon it by the detector is different from that for radiation passing through the filter. This means that it can't be flattened by a filtered flat-field exposure. If it is not subtracted off, then residuals in the response pattern remaining after normal flattening may degrade the quality of the image.

Thus, to carry out deep surface photometry in the NIR of a source whose extent is significant compared to the field of view, it is necessary to take special measures to characterize the background accurately. Normally, this means alternating between target and sky fields. The observing sequence, frequency of observing the sky, and the image reduction algorithm all have significant impacts upon the precision of photometry. Precisely how often the sky should be sampled is determined by the time scale and amplitude of background variations.

Published approaches to observing extended sources vary widely (e.g., Galaz 2000; Boselli et al. 1997; Gavazzi et al. 2001; Davidge 2000), and often appear to be based on personal experience or impressions. Few authors give hard evidence to back up their decisions. For example, on the question of how frequently to observe the background, we have heard advice ranging from 30% to 100% of the time spent on the target. Unfortunately, little quantitative information has been published about the speed, scale, and amplitude of background variations in standard astronomical bandpasses. Consequently, observers lacking experience in the near-infrared have very little concrete information to draw upon to decide the best approach to observe. Moreover, different observers use different methods to reduce their NIR images, methods which are based upon their own experience and which may be tied to their observing technique. Without knowledge of the optimal observing strategy, it is difficult to decide which approach is most appropriate.

We faced these issues recently when we were confronted with planning observations of faint dwarf irregular (dI) galaxies with the Canada-France-Hawaii Telescope (CFHT). Due to the sketchiness of published information and the wide scatter in advice from "experts", we were forced to adopt a strategy of maximum inefficiency, i.e., a conservative approach which would allow us to meet our minimum science goals while at the same time providing us with data suitable for discerning optimal strategies for the future. These data have enabled us to quantify NIR background variations, at least as seen by the CFHT-IR camera of CFHT atop Mauna Kea. Combined with experience gained from observing with another NIR array at the Observatorio Astronomico Nacional atop San Pedro Martir, Mexico, we have been able to develop guidelines for observing and reducing images of faint extended sources.

In § 2, the observations which form the foundation of our analyses are described. Temporal and spatial variations in the background are quantified in § 3. In § 4, we recommend some observing strategies, and in § 5 we describe the reduction method which yields the best results from our observations. Conclusions follow in § 6.

## 2. Observations

On three nights from 2002 Feb. 28 through Mar. 3, we used the CFHT-IR camera at the $f/8$ Cassegrain focus of the 3.6 m CFHT to observe faint dwarf galaxies in $J$ and $K'$. The CFHT-IR camera has at its heart a HgCdTe array with $1024 \times 1024$ pixels. The scale was $0\farcs211\,\mathrm{pix}^{-1}$, giving a field of view of $3\farcm6 \times 3\farcm6$. The gain was $2.3\,\mathrm{e}^-\,\mathrm{ADU}^{-1}$.

Although the allocated time was bright (1.8 to 4.2 days after Full Moon), the Moon affects observations by only about $0.9\,\mathrm{mag\,arcsec}^{-2}$ in $J$ and $0.3\,\mathrm{mag\,arcsec}^{-2}$ in $K$ (Vanzi & Hainaut 2002). Conditions were photometric on all three nights, with seeing $\sim 0\farcs5$, humidity less than 30%, and a steady temperature of about $-2°\,\mathrm{C}$ (UKIRT 2002), all typical of Mauna Kea when weather is good.

Many of our targets had apparent sizes which were more than 40% of the field of view of the camera. Therefore, we sampled the background by nodding the telescope by $3'$ to $5'$, i.e., by moving the target entirely out of the field. After starting with a sky frame, we alternated between target and sky, exposing each equally. To eliminate stars in the sky frames and to improve flattening, we adopted the classical approach of dithering each new sky frame and each new galaxy frame by $5''$ to $10''$ with respect to the previous sky or galaxy frame. Exposures in $J$ were restricted to 100 s and and those in $K'$ to 60 s. Between 25 and 30 s was lost between every pair of frames as a result of readout, nodding, and dithering. Over three nights, we acquired a total of about 200 images in $J$ and 600 images in $K'$.

Typically, near the zenith, the total count rate in $J$ was 40 ADU s$^{-1}$ and the count rate in $K'$ was 200 ADU s$^{-1}$. Signals were superimposed upon a pedestal of about 1200 ADU. Dark frames taken with a cold opaque filter revealed that the count rate due to background sources below the filter wheel was only 0.2 ADU s$^{-1}$. The count rate due to thermal sources located above the filter wheel was 11 ADU s$^{-1}$ in $J$ and 49 ADU s$^{-1}$ in $K'$ (see § 3.2).

Tracked but dithered exposures of the twilight sky were acquired for flat fielding. For each filter, a sequence of equal-duration exposures was taken as the twilight brightened or faded. How the signal in any given pixel correlated with the mean for all pixels was used to quantify the relative sensitivity of that pixel to light, and thereby generate a flat field frame. After eliminating bad pixels, all frames studied here were divided by a flat field frame created in this way (see § 4.4).

## 3.    Characteristics of the Background

### 3.1.    Overview

In principle, by observing the sky at different airmasses over a very short period of time (a few minutes), it is possible to separate the atmospheric contribution to the background from the thermal contribution by the dome, telescope, and camera (Wainscoat & Cowie 1992). Our data only allows us to study the stability of the combination of these two sources (to which we refer as "background"). However, we are able to conclude that changes in the background level arose primarily from variations in the brightness of the sky. As will be shown below, the background level correlated with celestial parameters, such as the airmass and the brightness of stars. Dark frames revealed that the contribution to the mean background level from below the filter wheel was small. Based upon the overall correlation of the background level with airmass, the thermal contribution in both $J$ and $K'$ was always less than 30% of the total (see § 3.2). Furthermore, the thermal background level should not have changed significantly from frame to frame because the outside temperature varied by at most 2 C over the entire run.

From 0.8 to 2.3 $\mu$m, the atmospheric background predominantly comes from airglow emission lines (e.g., Glass 1999). The lines originate in the high atmosphere (80 to 100 km altitude), primarily arising from transitions between vibrational states of the OH radical. All three NIR atmospheric windows accessible to ground-based observatories ($J$, $H$, and $K$) are particularly rich in OH emission lines. Due to changing atmospheric conditions, the strength of the OH bands is highly variable, so the NIR sky brightness is not constant in time or space. It has been reported to vary with season (Adams & Skrutskie 1997), with atmospheric temperature (Wainscoat & Cowie 1992), with airmass (Wainscoat & Cowie 1992; Adams & Skrutskie 1997), with time since sunset (Adams & Skrutskie 1997: the NIR sky is generally brightest in the first 4 to 5 hrs at the beginning of the night), and with location (Nguyen et al. 1996: at the South Pole, the sky in $J$ and $H$ is ~3 times darker than at Mauna Kea, and at $K$ it is 15 to 25 times darker). The surface brightness of the night sky can change by as much as 3 to 5% within 10 minutes (e.g., Galaz 2000; Gavazzi et al. 2001), and by up to 50% over the course of one night (Glass 1999).

Movies showing how the sky brightness at the zenith varies in $J$ and $H$ have been constructed by Adams & Skrutskie (1997) from sequences of images acquired with the 2MASS prototype camera. Over $9° \times 9°$, the airglow can be seen to vary dramatically, both temporally ($\sim 15\,\mathrm{s}$) and spatially. However, the images do not have sufficient resolution to reveal how the sky varies on a scale of a few arc minutes, which is what is relevant to imaging with near-IR detectors on large telescopes.

### 3.2. The Background Level

We observed, as have others, that the NIR sky brightness varies independently with the time since sunset and with the airmass. Superimposed upon those trends are random fluctuations, presumably caused by changes in the airglow amplitude and/or pattern or short-term variations in atmospheric transparency.

Figure 1 shows for both $J$ and $K'$ how the median of all pixels in each image, scaled to one second of exposure time, varied with the time of acquisition (UT) on the third night of our run. The data were derived from frames acquired both off and on the targets. Medians for on-target frames are not affected by target fluxes, because the targets were so faint relative to the background, and often significantly smaller than the field of view. Each observing sequence is labeled with a number, and the airmasses at the beginning and end are noted. Comparing sequences 1, 4, 5, 8 and 9, it is clear that the background at a given air mass declined steadily as the night progressed. As well, within each sequence, the background varied by an amount which constituted a significant fraction of the overall decline.

To better expose the variation with airmass, we modeled the trend in median surface brightness $I$ with observing time $t$ since sunset $t_0$ as

$$I = a + be^{-k(t-t_0)} \qquad (1)$$

The coefficients $b$ and $k$ define the rate of variation through the night, and $a$ is the asymptotic value of the count rate (the "zero-point"). We derived the free parameters using four sequences in $J$ and five sequences in $K'$ acquired on the third night near the zenith (airmasses between 1.0 and 1.1). The fits are shown as dotted curves in Figure 1. For the other two nights, the same values of $b$ and $k$ were adopted, and solutions for $a$ were derived. Each median for night $i$ was then corrected for time by subtracting $a_i - a_3 + be^{-k(t-t_0)}$, where $a_i$ is the zero-point for the night and $a_3$ is the zero-point for the third night.

Figure 2 presents for both bands the corrected median background level for each sequence (7 consecutive frames in $J$ and 21 in $K'$), normalized to one second, plotted as a function of airmass for all three nights. The sky brightness is expected to grow linearly with the airmass, because it is proportional to the column density of atoms and molecules contributing to the airglow (Wainscoat & Cowie 1992; Adams & Skrutskie 1997). Indeed, such a trend is seen in both $J$ and $K'$. Nevertheless, there are instances when this general dependence does not hold. The sequences marked by arrows in Figure 2 are observations taken on the same night over comparable ranges in airmass, but for which the median background signals differed significantly.

The contribution to the background from thermal sources can be estimated by extrapolating linearly the median count rate to a virtual airmass of zero (Wainscoat & Cowie 1992). From linear least squares fits to the data for the third night (dashed lines in Figure 2), the thermal background rate in the $J$ band at an airmass of zero was $11 \pm 3\,\mathrm{ADU\,s^{-1}}$, compared with an asymptotic value of $34\,\mathrm{ADU\,s^{-1}}$ for the total background rate at the zenith. In $K'$, the thermal background was $49 \pm 14\,\mathrm{ADU\,s^{-1}}$, compared with an asymptotic value of $172\,\mathrm{ADU\,s^{-1}}$ for the total background rate at the zenith. Thus, in both $J$ and $K'$, the atmosphere was responsible for *more* than 70% of the total background signal in each frame (the asymptotic values are lower limits to the actual backgrounds observed). Similar results were obtained for the other two nights and from a fit of all three nights together.

Table 1 summarizes the rates at which the background level (sky plus thermal) varied as judged from all of the sequences acquired during the run. Columns 1 and 2 give the night and the filter respectively, and columns 3 and 4 give the average and the maximum percentage deviations between the medians of successive pairs of frames, normalized to a time interval of one minute. From these data, it is possible to judge typical background level variations over one minute for a photometric night at Mauna Kea. In $J$, the average is 0.66% per minute and the maximum is 6.12% per minute. In $K'$, the average is 0.45% per minute and the maximum is 4.26% per minute. Based upon the estimated thermal background, variations relative to the sky alone are higher by less than 40%. One might expect more severe variations at wetter sites.

The potential precision of photometry depends upon whether variations are additive or multiplicative. If background variations are additive, as would be the case if only the atmospheric source function is changing, then the flux from a celestial source at a fixed airmass should be independent of the sky level. Photometric precision would not depend upon the amplitude of the variations. On the other hand, if background level variations are multiplicative, as they would be if they were linked to changing optical depth, then the flux from a celestial source would be correlated with the sky level. In this instance, photometric precision would be impaired as swings in surface brightness become more extreme. If the source function of the atmosphere is fixed, the rate of change of the magnitude $m$ of a celestial body with respect to the surface brightness $\mu$ of the sky should be given by

$$dm/d\mu = -kX/(2.5 \log e) \qquad (2)$$

where $k$ is the extinction coefficient, $X$ is the airmass, and $e$ is the base of natural logarithms.

We examined stars in several sequences of sky frames to elucidate the nature of the background level variations. Figure 3 shows for a typical sequence (spanning about 40 minutes) how the $K'$ magnitude of a star varied with the mean surface brightness of the sky (i.e., the background with the thermal component removed). Uncertainties in the measurements are smaller than the symbols. The sky magnitudes have been corrected for the long-term exponential decay with time using equation 1. The vertical bar displays the range of variation in the star's brightness expected from the changing altitude, based upon an extinction coefficient of 0.065 measured on the same night. The dotted line displays the correlation predicted by equation 2. Overall, the star's brightness faded as the sky brightened, proving that changes in sky brightness were accompanied by changes in atmospheric optical depth. However, the rate at which the star's brightness changed was much larger than simple radiative transfer predicts.

It is clear from Figure 3 that the signal from a celestial source is linked to the height of the pedestal upon which it stands. Consequently, photometry can be expected to be degraded if the amplitude of variation of the sky brightness is large, even if observing conditions appear to be photometric.

### 3.3. The Background Pattern

As Adams & Skrutskie (1997) have shown, the spatial distribution of the surface brightness of the night sky in the NIR is not uniform. Also, spatial variations in the background may occur as a result of instrumental effects, such as uneven cooling of the array or electronic interference. Variations in the electronics associated with each pixel can lead to large non-Poissonian fluctuations in signal with a "fixed" pattern. Instrumental signatures can be removed using sky frames taken at a rate frequent enough to overcome temporal variations. However, any atmospheric pattern present underneath an extended source cannot be removed this way. The best that can be done is to observe for a long enough time relative to the time scale for variation that deviations average out.

To examine the temporal stability of the spatial distribution of the background across the $3\farcm6$ field of view of CFHT-IR, we analyzed successive off-target and on-target pairs of frames in our observing sequences. The difference between the frames in each pair was computed after editing bad pixels, flattening, and additively adjusting them to the same midpoint. A median filter was applied to the difference image to reduce the noise relative to any residual pattern.

The panel in Figure 4 shows, in order of time, the smoothed differences between successive sky and target frames in one of the $K'$ sequences. The frames were taken during the course of observation of a faint galaxy only $1'$ in diameter. Differences between successive on-target frames are labeled with even numbers, and differences between successive off-target frames are labelled with odd numbers. All images in Figure 4 have been displayed with the same brightness and contrast. The differences are not flat and the pattern of deviations varies with time. Note that both target and sky sequences spanned about 40 minutes, with about 3 minutes between successive target or sky exposures.

The evolution of the patterns of Figure 4 is pretty continous, with patterns in on-target frames similar to patterns in off-target frames. This can be globally verified by comparing the average of the nine sky differences with the average of the nine target differences. The averages, which are displayed in Figure 5, show the same patterns. The images differ by less than 0.02% of the initial background level.

If the pattern were associated with the sky, it would appear different in images of the target, but this is not the case. It is concluded that the background pattern originated predominantly from the instrumentation rather than the sky. An independent analysis of successive pairs of dark frames revealed the same kind of patterns.

The amplitude of the pattern derived from each off-target image pair has been quantified by the difference between the most negative and positive deviations from zero. Table 2 lists the average and the maximum of these differences as a percentage of the median background level (sky plus thermal), normalized to a time interval of one minute. The figures were calculated using all sequences available for all three nights (about 280 pairs in $K'$ and 80 pairs in $J$). Relative to the total background level, the pattern varied by 0.34% in $J$ and 0.20% in $K'$ per minute on average. Based upon Figure 5, the rate of change of the spatial structure of the *atmospheric background* over the $3\rlap{.}'6 \times 3\rlap{.}'6$ field of view of CFHT-IR must have been less than 0.06% per minute in $K'$ on average relative to signal from the sky alone.

The changes that we observe in the background are not unique to our run. Archived data taken by independent observers about one year earlier (CADC 2003) using the same detector CFHT-IR, the same filters, and the same range of exposure times show similar variations in the background pattern and level. Such variations are also present in data acquired in 2002 by the first author using the Camilla $256 \times 256$ NICMOS3 array attached to the 2.1 m telescope at the Observatorio Astronomico Nacional in Mexico. Clearly, researchers must be vigilant in quantifying the rate and amplitude of background variations early on in order to judge how best to observe.

## 4. Recommendations for Observing

### 4.1. Exposure Times and the Background Sampling Frequency

Because of the brightness of the background, NIR exposure times are limited by the full-well capacity of pixels in the array. For example, the capacity of pixels in the CFHT-IR array is $45,000$ ADU, with response considered linear up to about $35,000$ ADU (CFHT 2002). During our run, the total background count rate peaked around 90 ADU s$^{-1}$ in $J$ and 390 ADU s$^{-1}$ in $K'$ . Thus, the maximum possible exposures were $\sim 390$ s and $\sim 90$ s in $J$ and $K'$, respectively. Whether or not exposures ought to be guided by these results depends upon the rapidity of background variations.

Because we lacked quantitative information about the stability of the sky and instrument, we decided to restrict exposure times in $J$ and $K'$ to 100 s and 60 s, respectively. Furthermore, target (T) and sky (S) fields were observed alternately, as follows:

$$S - T - S - T - S - \cdots - T - S \tag{3}$$

A sequence in $J$ consisted of 3 target frames and 4 sky frames. In $K'$, it was composed of 10 target frames and 11 sky frames. By using different subsets of the sky frames to reduce the target frames, it is possible to evaluate how the quality of reduced images depends upon the frequency with which the background is observed.

Figure 6 shows flattened background-subtracted images of Markarian 209 in $K'$ derived using successively fewer background samples. In all cases, the background in each target frame was subtracted using the nearest sky frame. The first panel shows the result when 10 of the sky images in the sequence are employed to reduce the 10 target images (effectively, as if the observing sequence were as above). The second, third, and fourth panels show results when 60%, 40%, and 20% as many sky images as target images are used, respectively, i.e., as if the observing sequences were as follows:

$$S - T - T - S - T - T - S - T - T - S - T - T - S - T - T - S \qquad (4)$$

$$S - T - T - T - S - T - T - T - S - T - T - T - T - S \qquad (5)$$

$$S - T - T - T - T - T - T - T - T - T - T - S \qquad (6)$$

To make imperfections clearer, the reduced images have been binned $5 \times 5$. The image derived using all sky samples is noticeably superior to the others. Significant blotching results when the sky is sampled less than 50% as often as the target, presumably because of the changing background pattern.

Clearly, in evaluating how long to expose, it is important not only to take into account how fast wells fill, but also the rapidity with which the background varies. The precision of background subtraction depends upon how much time separates the midpoint of a target exposure from the midpoint of a background exposure. To subtract the background *level* to a precision of 1% or better on average, midpoints should be separated by no more than 90 s in $J$ and 133 s in $K'$, given the average rates of change of the levels listed in Table 1 (0.66% in $J$ and 0.45% in $K'$ per minute). These are upper limits to the exposure times. In our case, about 10 s was taken up reading out the array and saving the data, and 15 to 20 s nodding from the target to the sky field, so exposures should not have exceeded 60 s in $J$ and 103 s in $K'$. With target and background exposures separated as described, the background *pattern* would be simultaneously removed with a precision of 0.5% in both $J$ and $K'$, given the average rate of change of the pattern listed in Table 2 (0.34% in $J$ and 0.20% in $K'$ per minute). If a separate background image is used with each target image, then any pedestal remaining in the average of background-subtracted target images should be significantly less than 1% of the original background.

It is important that the exposure time for background frames be the same as for target frames. This is because images contain instrumental signatures, like the bias, dark current, and non-Poissonian noise, which are not traced by the pattern in response. If the exposure for the background were different from that for the target, it would be necessary to multiplicatively scale the background frame by the ratio of exposure times to appropriately subtract the dark, thermal, and sky components from the target frame. However, this would mean that any bias or non-Poissonian noise would no longer be eliminated completely. Also, time-dependent instrumental signatures would not be removed as effectively.

## 4.2. Nodding

It is clear that the background underlying a target must be observed frequently. How it is observed depends upon the size of the field of view ($V$) relative to the diameter of the target ($D$).

If the target is small, with $V/D \gtrsim 3$, then it is possible to shift the pointing enough to sample the background under the entire target yet still keep the target on the array. If $V/D \gtrsim 4$, the target can be cycled through the centers of four quadrants of the array (by moving the telescope by $V/2$ in right ascension and/or declination between successive exposures). The target occupies any particular quadrant for fewer than half of the exposures, so an average background image free of contamination by the target can be created via median filtering (although, considering variability, this is not the best way to model the background– see § 5).

If the target is large, with $1 \lesssim V/D \lesssim 3$, then it is necessary to acquire background observations by moving the target completely off the array. To ensure that there is no chance of contamination by the target, any sky field should be at least $1.5V$ away from the target. Naturally, reductions are easier if the sky field has few bright stars or nonstellar objects in it.

## 4.3. Dithering

Sky fields are never blank. To make possible the removal of celestial contaminants, it is normal to shift the telescope between successive pointings at sky fields so that the contaminants move around on the array. If the field is uncrowded, then it is possible (although not necessarily desirable) to remove contaminants by median-filtering the background-subtracted target frames. If the field is crowded, then it is necessary to clean each sky frame before subtracting it from a target frame (see § 5).

Often, the dithering amplitude is set to be "several" times the FWHM of the point-spread function. We found this to be insufficient. Our frames were often contaminated by distant galaxies and/or bright stars. To remove stars well, sky frames should be dithered by at least 20 times the FWHM of the PSF ($10''$ in our case). To remove galaxies, shifts should amount to $20''$ or more.

It is also wise to dither images of target fields, but for a different reason. Once background-subtracted images are aligned, problems arising from imperfections in the array are removed by the averaging process. Also, errors in small-scale response corrections average out. In this case, the dithering amplitude need be only a few pixels.

### 4.4. Flat Fields

To correct for multiplicative variations in response across the field of view, it is necessary to acquire images of a uniform source of radiation whose optical path approximates that of celestial bodies. Ideally, the spectrum of the radiation ought to be similar to that of the predominant source of signal in images of astronomical targets, which would normally be the sky. In the NIR, the process is complicated by the presence of unwanted background contributions. We are aware of three methods for determining flat fields for NIR observations.

In the first method, the telescope is pointed at a surface inside the dome. An exposure is taken with the surface illuminated, and then an exposure of the same length is taken with the source of illumination turned off. The difference gives a true measure of pixel-to-pixel variations in response, because any bias, dark current and thermal background are removed. However, low frequency variations in response may not match those for celestial targets because of differences in the way the optics are illuminated. Furthermore, because the spectrum of illuminating radiation is very different from that .of the sky, the effective wavelength transmitted by a broad-band filter may differ significantly from that of the sky.

The second method is to observe the twilight sky. In this case, a sequence of dithered exposures of equal duration is taken as the twilight brightens or fades. The rate at which the signal in a pixel grows or fades relative to the average measures the relative response of that pixel, because contributions to the signal by the dark current and thermal background are the same for all exposures. The method has the advantage of accurately mapping low frequency variations in response if the telescope is baffled well. A disadvantage is that the timing of the observations is critical, and it may not be possible to acquire observations for more than two filters. Also, at twilight the spectrum of the sky is very different from that during the night, and is rapidly changing. Dithering ensures that each pixel is not exposed to a celestial source for the majority of the exposures.

A third method, which is employed often in the optical, is to create a so-called "superflat" by taking the median of sky frames acquired during the night (see Melnick et al. 1999). If the fields are not too crowded, and if positions are shifted sufficiently between exposures, then a flat field created in this way will be free of celestial contaminants. An advantage of this approach is that the background has a spectrum very close to that in each target field. Also, the illumination pattern should be close to that for the targets, unless there are significant position-dependent sources of scattered light (again, see Melnick et al. 1999). In the NIR, though, the approach is compromised by background radiation coming from sources other than the sky. As will be shown below, the median of sky frames acquired in any given sequence during our run was degraded by the changing spatial pattern of the instrumental background. Furthermore, there is the possibility that the night sky will not be an even source of illumination due to structure in the airglow.

For our run, we employed the second method. The relative response of each pixel was derived by finding the slope of the relation between the signal in the pixel and the median signal for the whole array. Computations were done with an IDL script created by Lai (2002), modified to use a robust least absolute deviation linear fitting technique (i.e., the LADFIT function of IDL) to avoid perturbations by stars still visible in the twilight. The grid of calculated slopes became the flat field image.

In acquiring twilight flats, one must be careful to observe during a small window of time when the sky is

1. bright enough that the exposure sequence can be completed in a time span of a few minutes. Over a longer period, a growing thermal background may compromise the derivation of the flat field.

2. faint enough to avoid unacceptably short exposures. With exposures of 0.1 seconds or less, a shutter can leave a pattern in the flat field because different parts of the field are illuminated for different lengths of time. Also, uncertainties in the timing may lead to more significant variations in the dark and thermal contributions to the background, which must be assumed to be fixed when the flat field is derived.

3. varying rapidly enough that the signal due to the sky changes significantly (by at least a factor of two) over the period of observation.

A sequence of exposures of fixed duration was started when the sky brightness was about 10 times higher than in darkness, which at the zenith occurred about 10 minutes in $J$ and about 2 minutes in $K'$ after sunset or before sunrise. Exposures were 2 to 3 seconds long. Frames were acquired (usually 20 at a time) until the sky level was a factor of about five higher than the starting value but well within the linearity limit. The sky field was tracked to avoid star trails, and the pointing was dithered to ensure that all pixels were free of stars for the majority of the exposures.

## 5.  Recommendations for Reductions

Suppose that a sequence of images of an extended target and adjacent sky fields is acquired. Consider a single image of the target, $T_i$. A flattened background-free image, $T_i^0$, is obtained from

$$T_i^0 = \frac{T_i - S_i'''}{F} \tag{7}$$

where $S_i'''$ (the notation is explained below) is an appropriately tuned sky frame with the same effective exposure as $T_i$, and $F$ is the flat field. A final reduced image of the target, $< T_i^0 >$, is obtained by aligning and combining all $T_i^0$.

The main unknown in the formula is $S_i'''$. $S_i'''$ must be chosen in such a way that the background in $< T_i^0 >$ ends up being flat and as close to zero as possible across the whole image.

$S_i'''$ has to be constructed using one or more images of the sky $S_i$ taken at different times and locations. It might be obtained in the following ways:

1. From the nearest sky frame. If exposure times are short enough, then the level and pattern of the background should be reasonably close to those in the target frame. Celestial contaminants in $S_i'''$ will lead to negative residuals in $T_i^0$, but if the field is not too crowded, these will not appear in $< T_i^0 >$ if it is computed as the median of all $T_i^0$.

2. From two sky frames (presumed dithered) taken before and after the target frame. The advantage of having two frames straddling the target frame is that the background at the time of observation of the target can be interpolated. Also, contaminants can be readily recognized and removed from each $S_i$ (see below), and the signal-to-noise ratio in $< T_i^0 >$ can be optimized by arithmetically averaging all $T_i^0$. The disadvantage is that the process of cleaning individual sky frames can be time-consuming.

3. From all sky frames in the sequence. The complicating factor is the changing sky level. If $S_i'''$ were computed as the median of all $S_i$ without compensating first for the varying level, then celestial contaminants would not necessarily be removed and would propagate into $< T_i^0 >$. If the target is small enough that all $S_i$ (and $T_i$) can be equalized (additively) to a common background level, then the median ought to be free of celestial contaminants, provided that the dithering amplitude was sufficiently large. If the mean time of observation of the sky frames coincides with that of the target frames, then any pattern in $S_i'''$ ought to be similar to that in $< T_i >$, and thereby not appear in $< T_i^0 >$.

We experimented intensively with all three methods of computing $S_i'''$. The second yielded the finest results for $< T_i^0 >$. Figure 7 shows the reduced images for Markarian 209, binned $5 \times 5$, which result from the second and third methods. The image resulting from the first method is presented in Figure 6a. The background resulting from the second method (Figure 7a) is smooth. Blotchiness is apparent in the background of the image derived from the third method (Figure 7b), probably because the effective time of acquisition of the median was not quite the same as the mean time at which the frames of the galaxy were acquired. As can be seen from Figure 6a, the first method gives a result similar to the second method. However, the signal-to-noise ratio in the unbinned image is inferior, probably because only one sky frame, as against the average of two, was used to estimate the background in each target frame. For large galaxies, we found that the signal-to-noise ratio and the flatness of the residual background in $< T_i^0 >$ steadily decreased as the number of sky frames utilized to compute $S_i'''$ increased above two. Note that the third method is unusable if the target is close to filling the field of view.

Given the following observing sequence,

$$S_1 - T_1 - S_2 - T_2 - \cdots - S_i - T_i - S_{i+1} - \cdots - S_n - T_n - S_{n+1} \tag{8}$$

and given that sky and target exposures are identical, we recommend the following reduction procedure:

1. Map out bad pixels, and correct them in all frames.

2. Create a flat field image, and divide it into all $S_i$ and $T_i$. Even though there are components in the background which do not have the response pattern superimposed, taking this step at the beginning allows additive adjustments to the images to be made with impunity. Since the duration of exposures is fixed, residuals introduced by flat-fielding these components ultimately subtract out.

3. For each $T_i$, compute the difference $(\Delta S)_i$ between straddling sky frames:

$$(\Delta S)_i = S_i - S_{i+1}$$

   Celestial sources formerly lost in the non-Poissonian noise of the individual sky exposures will appear as positive and negative pairs (displaced by dithering).

4. Find all of the positive sources in $(\Delta S)_i$, say by using DAOFIND in IRAF.

5. Remove the positive sources in $(\Delta S)_i$, say by using IMEDIT in IRAF non-interactively.

6. Create a contaminant-free sky image $S_i'$ by adding back $S_{i+1}$:

$$S_i' = (\Delta S)_i + S_{i+1}$$

7. Switch $S_i$ and $S_{i+1}$, and repeat the four steps above to derive $S_{i+1}'$.

8. Average or interpolate (in time) $S_i'$ and $S_{i+1}'$ to compute $S_i''$, the best approximation to the background at the time of observation of $T_i$. For example,

$$S_i'' = \left(S_i' + S_{i+1}'\right)/2$$

9. If the object is small enough, measure the background levels $T_{i,b}$ and $S_{i,b}''$ in the corners of $T_i$ and $S_i''$, respectively. Then compute the adjusted sky image $S_i'''$:

$$S_i''' = S_i'' + \left[T_{i,b} - S_{i,b}''\right]$$

The background level across $S_i'''$ will closely match that across $T_i$. If the object is close to filling the field, then $S_i'''$ must be taken to be $S_i''$.

10. Subtract off the background from $T_i$ to create a background-free image of the target, $T_i^0$:

$$T_i^0 = T_i - S_i'''$$

11. Repeat all steps above for each $T_i$.

12. Align all $T_i^0$ and, if there are enough images, take the median rather than the average (with rejection) to eliminate any remaining transients, such as cosmic rays. An arithmetic average leads to a higher signal-to-noise ratio, but if any stars or galaxies were not removed from the sky frames, say because DAOFIND didn't find them, then the final image will be compromised by their residuals. Additive leveling, either during the reductions or when combining, aids in the rejection of the faintest transients.

The procedure above led us to images with a background level typically less than 0.1% of that in the original frames. Low-frequency variations in the background were imperceptible, implying that their amplitude was less than about 0.03% of the signal in the original frames.

## 6. Conclusions

To plan observations of extended sources in the NIR, it is important to have quantitative information on how the background varies in space and in time. Especially, how fast the background changes affects choices for exposure times, observing sequences, and reduction strategies. Using observations acquired with an IR array at CFHT during three photometric nights, we have studied changes in the background level and pattern in $J$ and $K'$, and arrived at recommendations for approaches to observing and reductions.

In observing each target, the background was sampled as often as the target by alternating between target and sky fields. The exposure time for each frame was only 100 s in $J$ and 60 s in $K'$.

The majority of the background (more than 70%) was contributed by the sky. The level varied significantly on a time scale of a minute, but also tended to rise with increasing airmass and decline with the time since sunset. Short-term variations amounted to 0.7% per minute in $J$ and 0.5% per minute in $K'$, on average, relative to the total background signal. As the sky brightened, stars faded, showing that variations in the level of the sky were accompanied by changes in optical depth.

A temporally variable pattern in the background was revealed by subtracting consecutive images of the sky. Subtraction of consecutive target frames revealed the same pattern, so its origin lay with the instrument, not the sky. The amplitude of the pattern changed at a rate of 0.34% per minute in $J$ and 0.20% per minute in $K'$ on average relative to the total background signal. In $K'$, any pattern in the atmospheric background varied by at most 0.06% per minute in $K'$ relative to the signal from the sky alone.

Separate observations reveal that the background variations described here were neither unique to our run nor our instrument. Analysis of a different set of data taken at CFHT by independent researchers about one year earlier led to comparable results for the variation of the background level and its pattern. Data taken by us with a different array at a site in Mexico also showed similar problems.

Rates of variation of the level and pattern of the sky lead to stringent constraints on exposure times and background sampling frequencies. It is imperative to alternate between target and sky positions. Exposure times for individual frames of target and sky fields ought to be identical. To enable subtraction of the background to 1% accuracy, target and background exposures should be separated by no more than 90 s in $J$ and 130 s in $K'$ (at least, at Mauna Kea). If the target size exceeds about 40% of the field of view, the target ought to be moved completely off the array to sample the background. If the target is smaller, it is possible to increase observing efficiency by keeping the target on the array during background observations.

Good flat field images can be acquired by taking a sequence of short equal-duration exposures as the twilight sky fades or brightens. The rate at which the signal in a pixel correlates with the mean signal in a frame gives the response of that pixel to the *varying* component of illumination (the sky), relative to the mean.

It is prudent to begin reductions by flattening all frames. In so doing, it becomes possible to make additive adjustments to background frames to level them with target frames if targets don't fill the field.

The best approach to removing the background in alternated target and background frames is to solve for the background of each target frame individually. This can be done by averaging or interpolating background frames taken on either side of a target frame. Celestial contaminants can be recognized and removed by first subtracting the two background frames from each other. For smaller sources, the background level in the average can be matched to that in the target frame by measuring the signal in the corners.

Although our data about background rates are specific to CFHT, conclusions about observing and reduction strategies should have wider value. Certainly, we have shown that it is imperative for any NIR observer of extended sources to have quantitative knowledge of the rate of variation of the level and pattern of the background prior to observing anywhere. Armed with such data, it is hoped that the information in this paper will aid in the development of effective observing and reduction plans.

We thank the CFHT time allocation committee for granting us the opportunity to observe. CFHT is operated by the National Research Council of Canada, the Centre National de la Recherche Scientifique, and the University of Hawaii. MLM is grateful to the Natural Sciences and Engineering Research Council of Canada for its continuing support. Thanks are conveyed to the National Research Council of Canada for funding the observing expenses of OV. For our data reductions, we used IRAF, distributed by the National Optical Astronomy Observatories, which are operated by the Association of Universities for Research in Astronomy, Inc., under cooperative agreement with the National Science Foundation. To compare additional data with our own, OV registered as a Guest User of the Canadian Astronomy Data Centre, which is operated by the Dominion Astrophysical Observatory for the National Research Council of Canada's Herzberg Institute of Astrophysics. Special thanks are conveyed also to the anonymous referee whose suggestions helped us to improve the paper.

## REFERENCES

Adams, J. D., & Skrutskie, M.F. *Airglow and 2MASS Survey Strategy*, on-line at http://pegasus.phast.umass.edu/adams/air.ps

Boselli, A., et al. 1997, A&AS, 121, 507

CFHT (Canada-France-Hawaii Telescope) 2002, at http://cfht.hawaii.edu

CADC (Canadian Astronomy Data Centre) 2003, CFHT archive at http://cadcwww.dao.nrc.ca/cfht/

Davidge, T. J. 2000, AJ, 119, 748

Fioc, M., & Rocca-Volmerange, R., 1999, A&A, 351, 869

Galaz, G. 2000 AJ, 119, 2118

Gavazzi, G., et al. 2001, A&A, 372, 29

Glass, I. S., 1999, *Handbook of Infrared Astronomy* (Cambridge: Cambridge University Press)

Lai, O. 2002, personal communication

Melnick, J., Selman, F., & Quintana, H. 1999, PASP, 111, 1444

Nguyen, H. T., et al. 1996, PASP, 108, 718

UKIRT (United Kingsdom Infrared Telescope) 2002, Graphical Weather Server on-line at http://www.ukirt.jach.hawaii.edu/weather_server/weather.cgi

Vanzi, L., & Hainaut, O. R. 2002, on-line at European Southern Observatory, http://www.eso.org/gen-fac/pubs/astclim/lasilla/l-vanzi-poster

Wainscoat, R. J., & Cowie, L. L. 1992, AJ, 103, 1, 332

Table 1.   Percentage Change in NIR Background Level Per Minute

| Night | Filter | Average | Maximum |
| --- | --- | --- | --- |
| 1 | J | 0.46 | 2.23 |
| 2 | J | 0.73 | 6.12 |
| 3 | J | 0.78 | 4.20 |
| 1 | $K'$ | 0.37 | 2.34 |
| 2 | $K'$ | 0.57 | 4.26 |
| 3 | $K'$ | 0.41 | 3.64 |

Table 2.   Percentage Change in NIR Background Pattern per Minute

| Night | Filter | Average | Maximum |
|-------|--------|---------|---------|
| 1 | J | 0.35 | 0.92 |
| 2 | J | 0.31 | 0.92 |
| 3 | J | 0.35 | 0.91 |
| 1 | $K'$ | 0.19 | 0.48 |
| 2 | $K'$ | 0.20 | 0.43 |
| 3 | $K'$ | 0.22 | 0.65 |

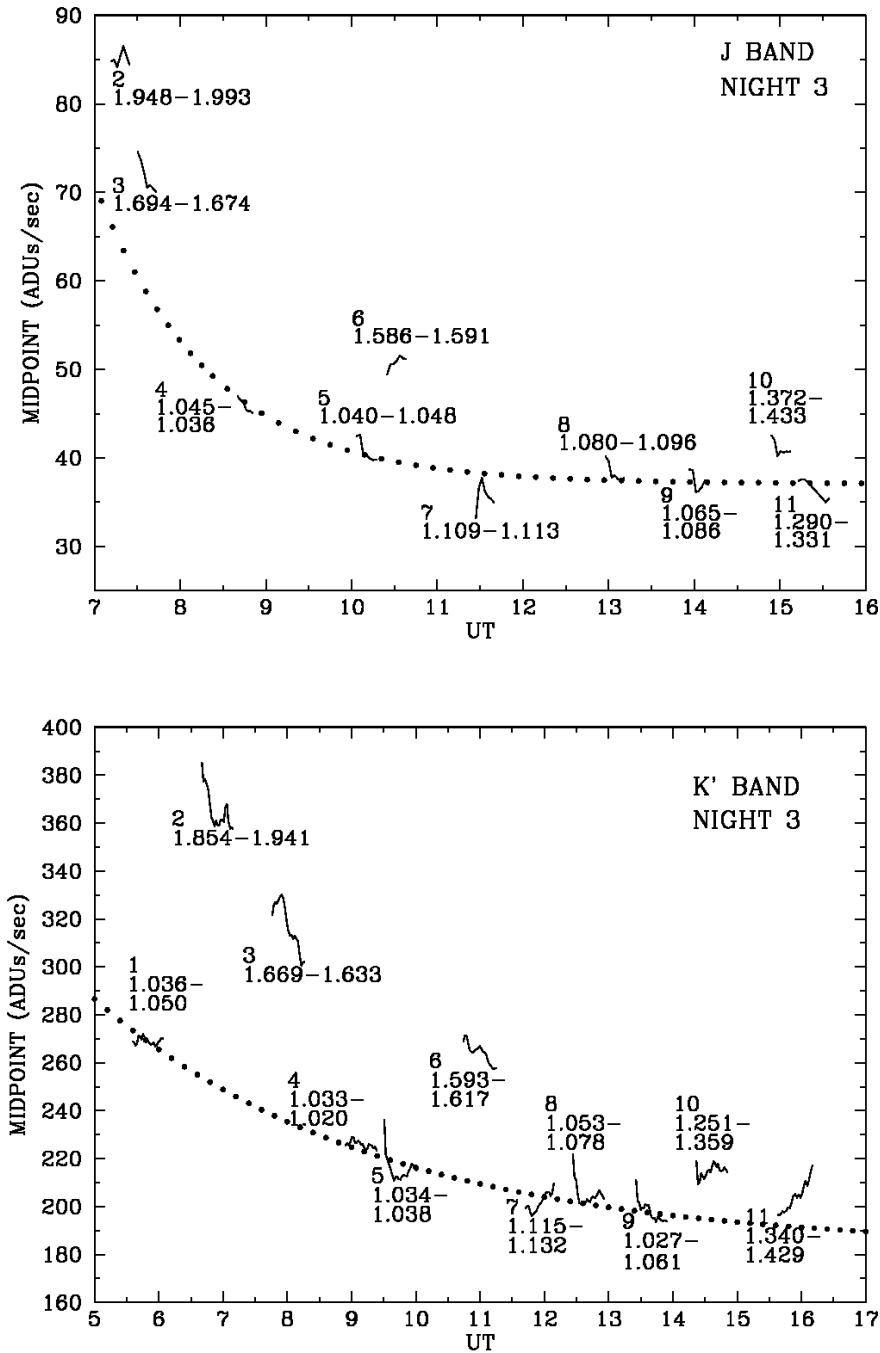

Fig. 1.— Median background level (scaled to one second) for $J$ (top) and $K'$ (bottom) as a function of Universal Time on the third night. Each sequence is labeled with a number and the range of airmass covered. Even ignoring sequences 2 and 3, whose backgrounds are enhanced by their high airmasses, there is a tendency for the level to decline as the night progresses. The dotted curves show the models fitted to data with air masses between 1.0 and 1.1.

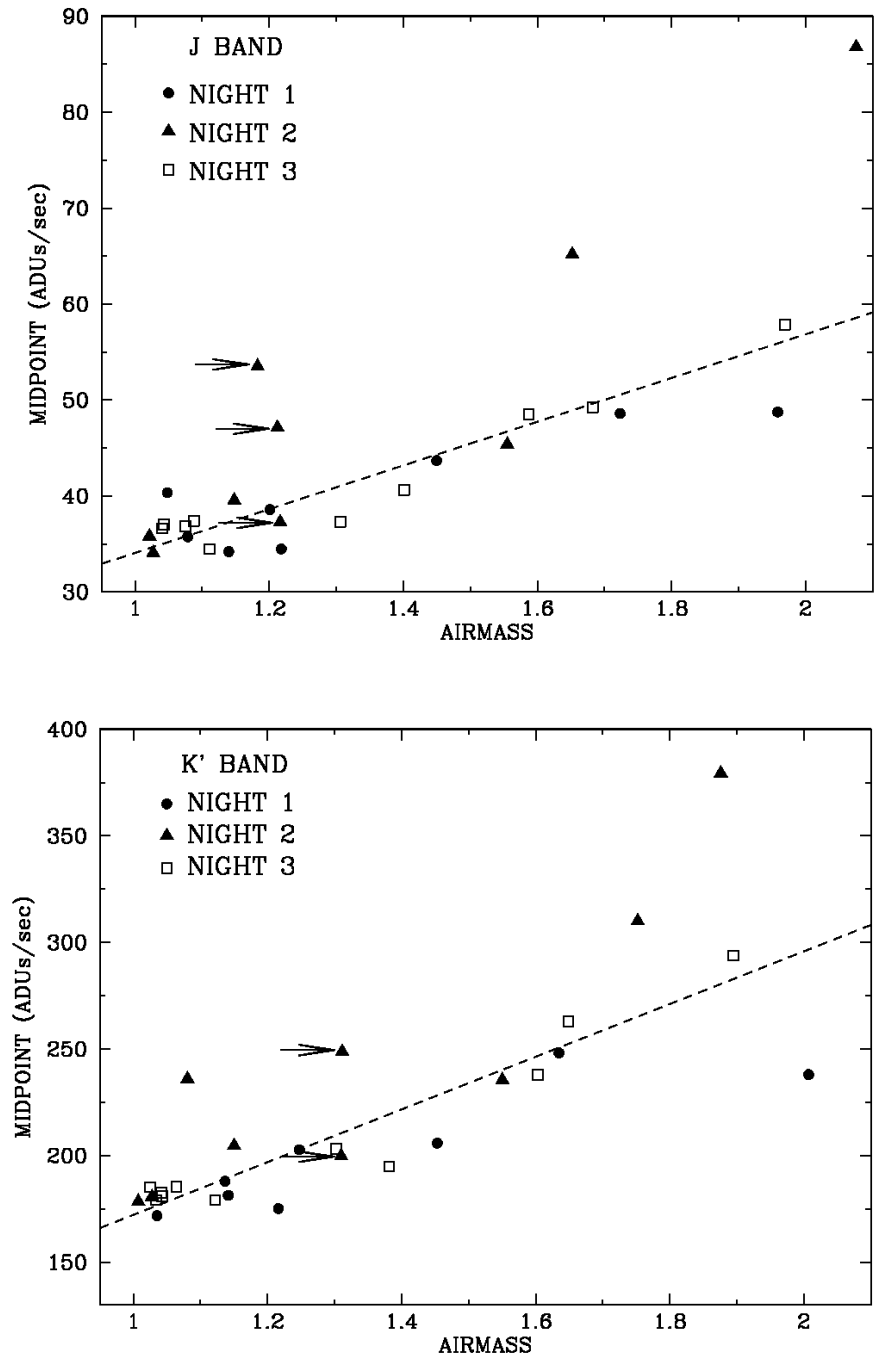

Fig. 2.— Temporally-corrected median background level (scaled to one second) for $J$ (top) and $K'$ (bottom) as a function of airmass for all frames in all observing sequences over 3 nights. Although there is an overall increase in the background with the airmass, some sequences at the same airmass, such as those marked by arrows, show different background levels. The dashed lines are linear least squares fits to the data acquired on the third night.

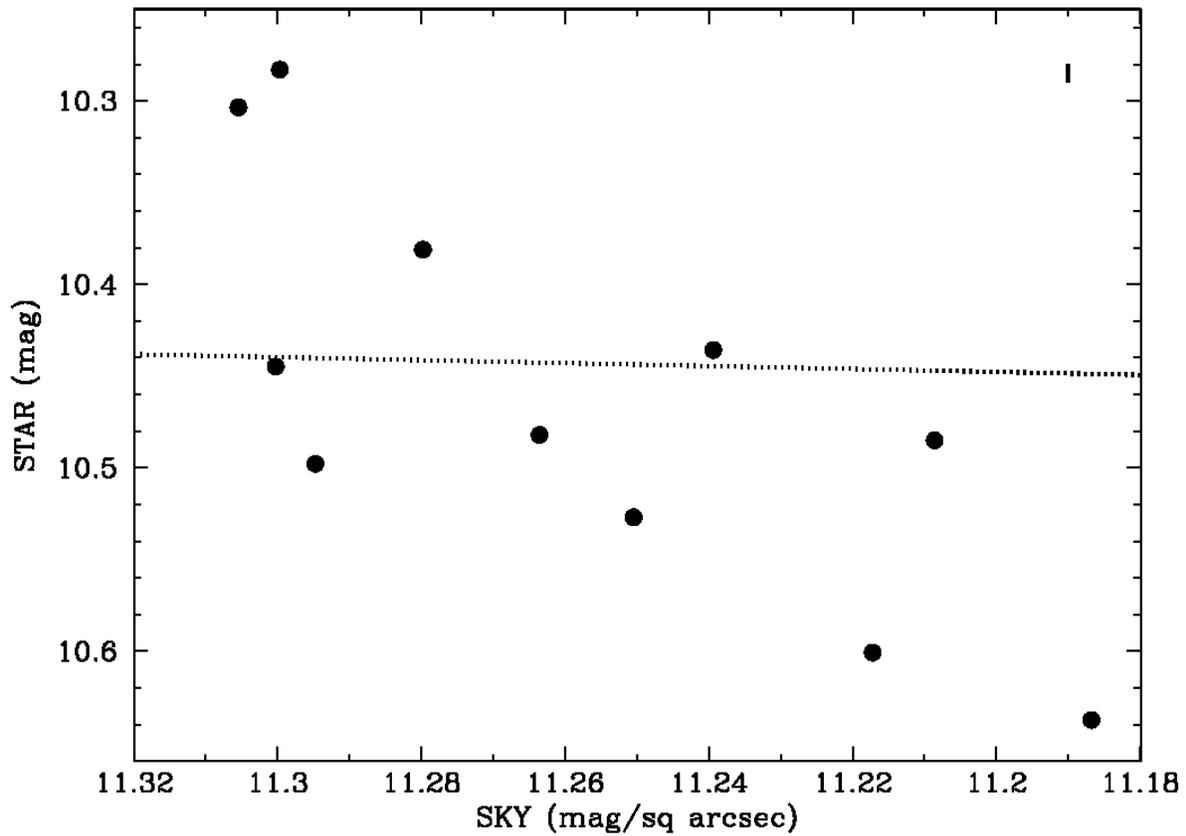

Fig. 3.— The $K'$ magnitude of a star as a function of the mean surface brightness of the sky in $K'$ (thermal background removed). Observations span about one hour, and uncertainties are smaller than the symbols. The vertical bar in the upper right corner shows the range of variation in the brightness of the star expected as a result of the changing altitude. The star clearly faded as the sky brightened, but not because of the altitude variation. The slope of the dotted line conveys how fast the star would be expected to fade if variations in the sky brightness were caused by fluctuations in optical depth alone. The intercept of the line has been set arbitrarily.

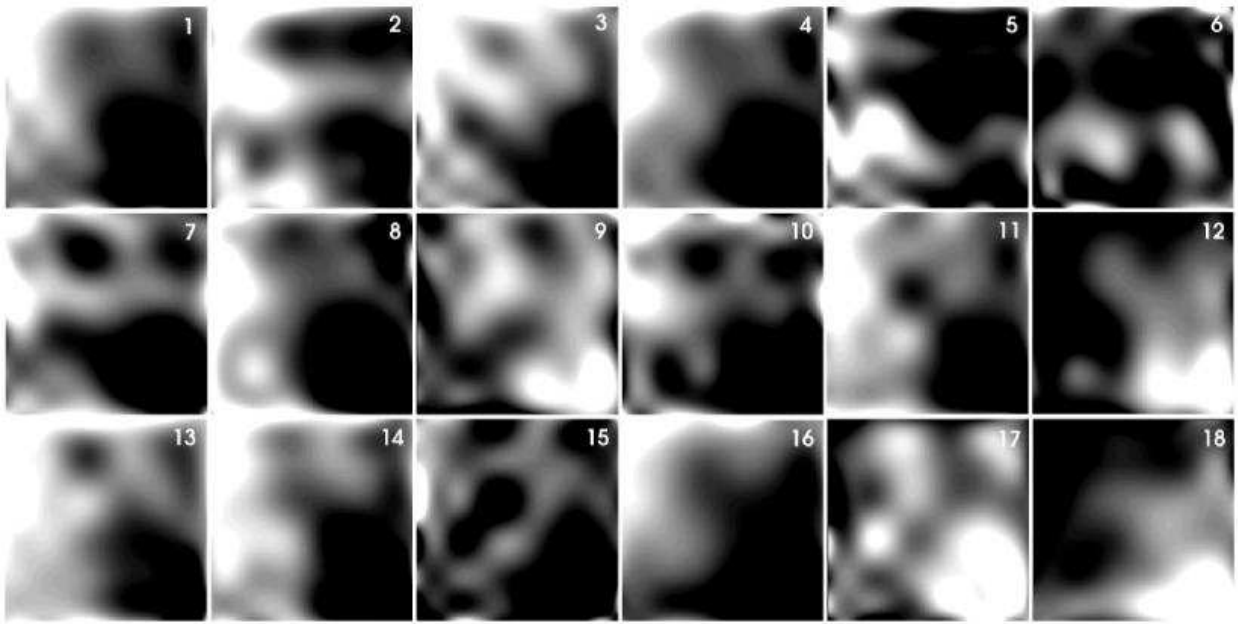

Fig. 4.— Mosaic showing the smoothed differences between successive sky frames (odd numbers) and successive target frames (even numbers) in a single $K'$ observing sequence spanning one hour. Variations amount to 0.2% of the median background signal per minute. The continuity between sky and target frames suggests that the pattern originates in the instrument, not the sky.

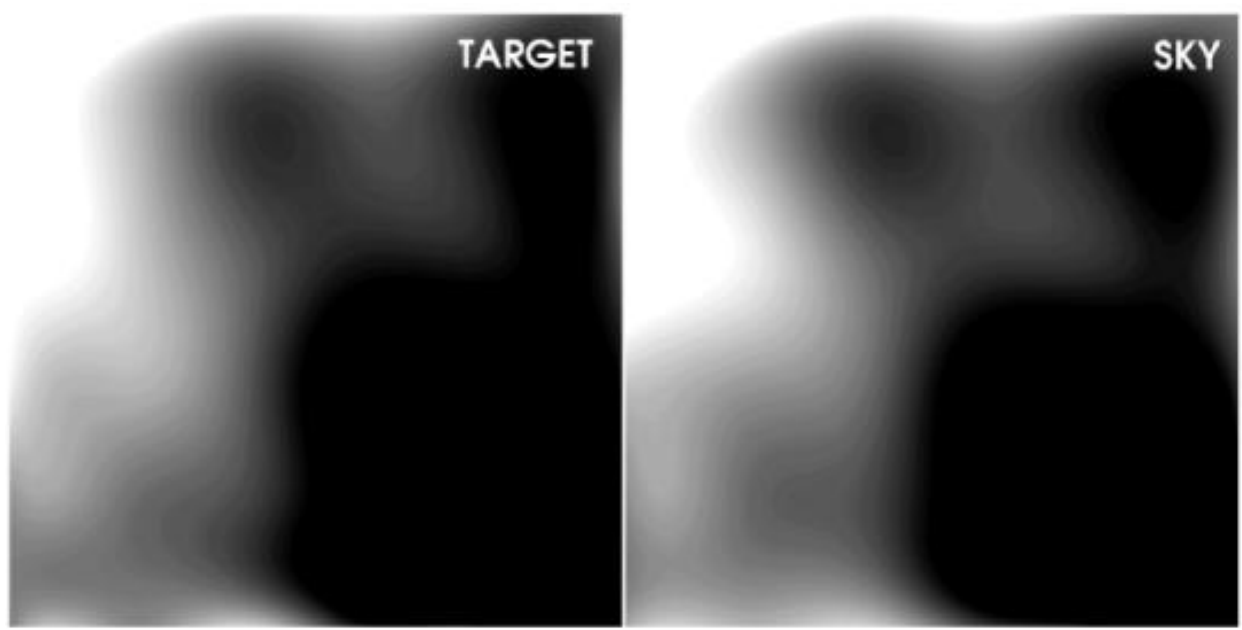

Fig. 5.— Averages of the smoothed differences between successive sky frames and successive target frames in a single $K'$ observing sequence spanning one hour. The images look similar, implying that the origin of pattern variations lies in the instrumentation rather than the airglow.

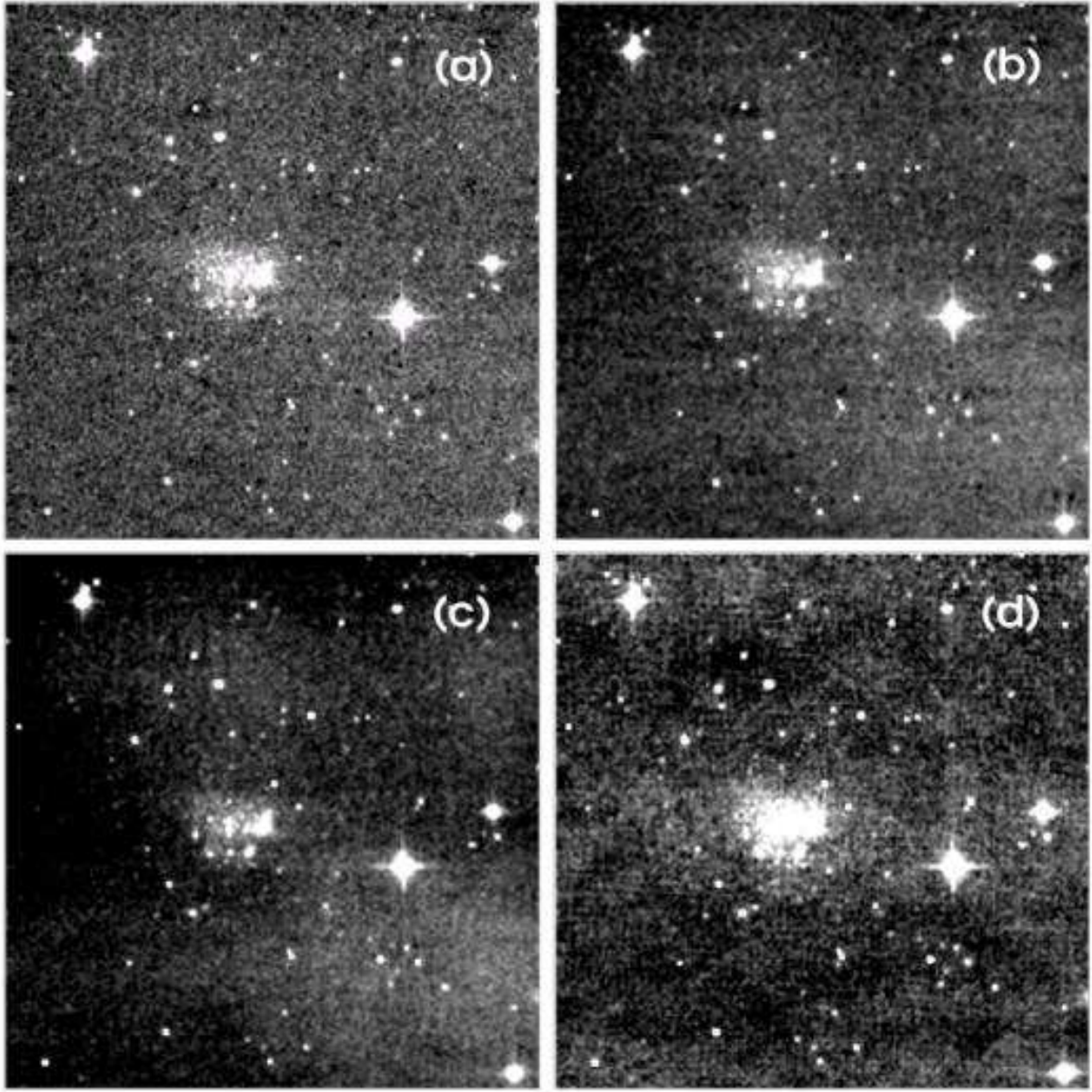

Fig. 6.— Reduced images of Markarian 209 constructed from a single sequence of 10 galaxy frames alternated with 11 sky frames, but with the background sampled four different ways. All are binned 5 × 5 and displayed using the same settings for contrast and brightness. The number of sky frames employed in creating the images was (a) 10, (b) 6, (c), 4, and (d) 2. Note how the residual background becomes more blotchy as the sky is sampled less frequently.

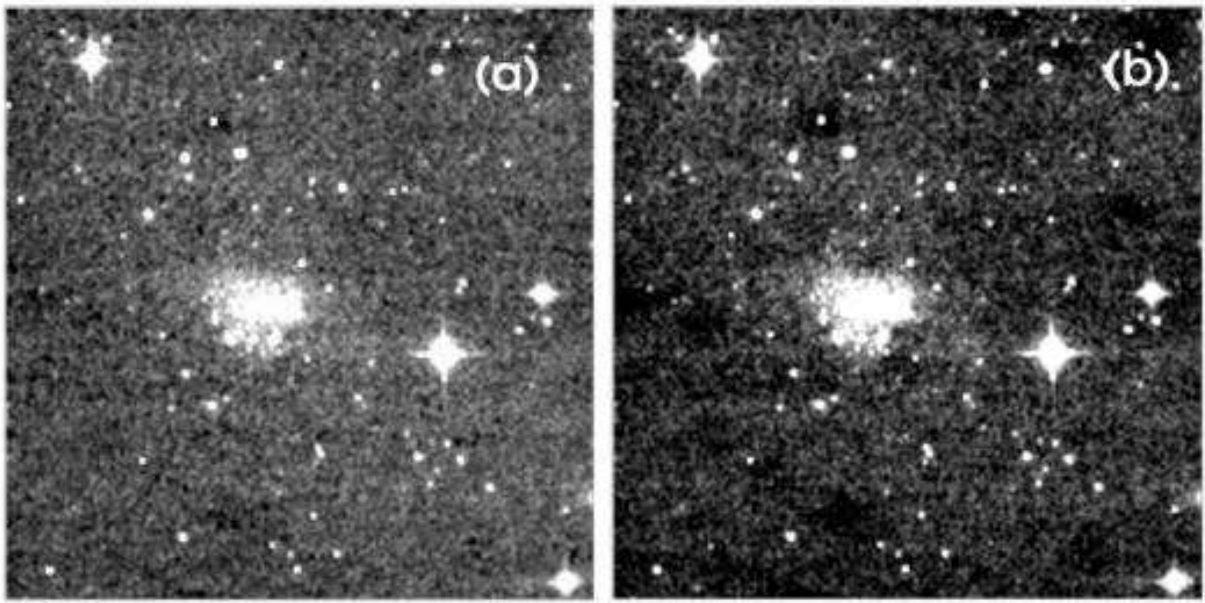

Fig. 7.— Reduced images of Markarian 209 constructed by subtracting from each galaxy frame (a) the average of the two sky frames taken immediately before and after, and (b) the median of all sky frames. Binning and contrast and brightness settings are the same as in Figure 6. The background is less smooth in the image derived from the sky median.